\documentclass[prd,aps,nofootinbib,showpacs,showkeys,preprintnumbers]
{revtex4}
\usepackage{graphicx,epsf,amsmath,amsfonts,amssymb,amsbsy}
\usepackage{epsfig}
\newcommand{\ds}{\displaystyle}
\newcommand{\vev}[1]{\langle#1\rangle}
\newcommand{\mat}{\left ( \begin{array}}
\newcommand{\emat}{\end{array} \right )}
\newcommand{\vect}{\left ( \begin{array}{c}}
\newcommand{\evect}{\end{array} \right )}

\newcommand{\Det}{\mathop{\rm Det}\nolimits}
\preprint{HU-EP-08/14}

\begin{document}
\title{ \bf Finite size effects in the
Gross-Neveu model with isospin chemical potential}
\author{D.~Ebert$^1$, K.G.~Klimenko$^{2,3}$, A.V.~Tyukov$^4$, and
V.Ch. Zhukovsky$^{4}$ }
\affiliation{$^1$ Institute of Physics, Humboldt-University Berlin, 
12489 Berlin, Germany}
\affiliation{$^2$ Institute
for High Energy Physics, 142281 Protvino, Moscow Region, Russia}
\affiliation{$^{3}$ Dubna University (Protvino
branch), 142281 Protvino, Moscow Region, Russia}
\affiliation{$^4$ Faculty of Physics, Moscow State University, 119991
 Moscow, Russia}

\begin{abstract}
The properties of the two-flavored Gross-Neveu model in the
(1+1)-dimensional $R^1\times S^1$ spacetime with compactified space
coordinate are investigated in the presence of the isospin chemical
potential $\mu_I$. The consideration is performed in the limit
$N_c\to\infty$, i.e. in the case with infinite number of colored
quarks. It is shown that at $L=\infty$ ($L$ is the length of the
circumference $S^1$) the pion condensation phase is realized for
arbitrary small nonzero $\mu_I$. At finite values of $L$, the phase
portraits of the model in terms of parameters $\nu\sim\mu_I$ and
$\lambda\sim 1/L$ are obtained both for periodic and antiperiodic
boundary conditions of the quark field.
It turns out that in the plane $(\lambda,\nu)$ there is a strip
$0\le\lambda<\lambda_c$ which lies as a whole inside the pion
condensed phase. In this phase the pion condensation gap is an
oscillating function vs both $\lambda$ (at fixed $\nu$) and $\nu$
(at fixed $\lambda$).
\end{abstract}


\maketitle

\section{Introduction}

It is well-known that QCD is a fundamental theory of strong
interactions both in the vacuum and in hot and/or dense baryonic
matter. However, it can be successfully used only in the region of
high energies, temperatures and densities (or chemical potentials),
where a weak-coupling expansion is applicable. Away from this
region, different nonperturbative methods or effective theories such
as chiral effective Lagrangians as well as Nambu -- Jona-Lasinio
type models (see, e.g., the papers
\cite{ebert_volkov,volkov,hatsuda,alford} and references
therein), are usually employed for the consideration of light meson
physics, phase transitions in dense quark matter, etc. In
particular, motivated by the fact that in heavy-ion collisions and
compact stars the hadronic matter is isotopically asymmetric,
different QCD-like effective models were studied at nonzero isospin
chemical potential $\mu_I$ \cite{son,frank,ek,andersen,abuki}. There
the charged pion condensation phenomenon, which is generated if
$\mu_I$ is greater than the pion mass $m_{\pi}$, was also
considered.

In all above mentioned papers the effective models are i) field
theories in usual (3+1)-dimensional spacetime, and ii) they are
employed for the description of QCD at {\it rather low} energies and
densities. At the same time there is another class of theories that
can be used as a laboratory for a qualitative consideration of QCD
at {\it arbitrary} energies. These are the so-called Gross-Neveu
(GN) type models, i.e. two-dimensional quantum field theories with
four-fermion interactions \cite{gn,VSMT,ft}. Renormalizability,
asymptotic freedom as well as the spontaneous breaking of chiral
symmetry (in the vacuum) are the most fundamental features that are
inherent both for QCD and all GN type models. In addition, the GN
phase portrait in terms of baryon chemical potential $\mu_B$ vs
temperature resembles qualitatively to a great extent the QCD phase
diagram \cite{wolff,kgk1,barducci,chodos,thies}. Due to their
relative simplicity in the leading order of a large $N_c$-expansion
($N_c$ is a number of colored quarks), it is very convenient to use
GN models for considering such a phenomenon of dense QCD as color
superconductivity \cite{chodos,zhou} and to elaborate new
nonperturbative methods of quantum field theory
\cite{okopinska,osipov,kgk}. Moreover, the influence of the space
compactification on chiral symmetry breaking both in the vacuum
($\mu_B =0$) \cite{kim} and in dense baryon matter ($\mu_B\ne 0$)
\cite{vshivtsev} was studied in terms of GN models (see also the
appropriate papers \cite{beneventano,gamayun,abreu}).

Before investigating different physical effects relevant to a real
(3+1)-dimensional world in the framework of two-dimensional GN
models, let us recall that there is a  no-go theorem forbidding the
spontaneous breaking of continuous symmetries in two dimensions
\cite{coleman}. However, at present time it is well understood (see,
e.g., the discussion in \cite{VSMT,barducci,chodos,thies}) that in
the limit $N_c\to \infty$ this no-go theorem does not 
apply. This makes it possible to study symmetry breaking effects in
terms of GN models as well, but only in the leading order of the
$1/N_c$-expansion, where most low dimensional theories are exactly
solvable. In this sense, for $N_c\to \infty$, low dimensional
quark models are physically more tractable and appealing than at
finite $N_c$.

In the present paper the pion condensation phenomenon is investigated
in the framework of the two-dimensional GN model with two massless
quark flavors. In particular, we shall study the influence of the
finiteness of the system size on this phenomenon. So our
consideration is performed in a spacetime with non-trivial topology,
i.e. on the $R^1\times S^1$ manifold with compactified space
coordinate, and the GN model is extended by an isospin chemical
potential $\mu_I$ (for simplicity, we put $\mu_B =0$). Obviously, the
latter issue is motivated by the physics of compact stars, where pion
condensation might be realized as a consequence of the isotopic
asymmetry of baryon matter. Since all the calculations are carried
out on the basis of the leading order of $1/N_c$-expansion 
(i.e. in the case $N_c\to\infty$) we expect that all conclusions
concerning the pion condensation phenomenon, caused by a spontaneous
breaking of the continuous isospin symmetry, remain qualitatively
valid for real QCD.

\section{The case of $R^1\times R^1$ spacetime}
\subsection{ The model and its thermodynamic potential}

We consider a two-dimensional model which describes dense quark
matter with two massless quark flavors ($u$ and $d$ quarks). 
Its Lagrangian has the form
\begin{eqnarray}
&&  L=\bar q\Big [\gamma^\nu\mathrm{i}\partial_\nu
+\frac{\mu_B}{3}\gamma^0+\frac{\mu_I}2 \tau_3\gamma^0\Big ]q+ \frac
{G}{N_c}\Big [(\bar qq)^2+(\bar q\mathrm{i}\gamma^5\vec\tau q)^2 \Big
],  \label{1}
\end{eqnarray}
where the quark field $q(x)\equiv q_{i\alpha}(x)$ is a flavor doublet
($i=1,2$ or $i=u,d$) and color $N_c$-plet ($\alpha=1,...,N_c$) as
well as a two-component Dirac spinor (the summation in (\ref{1})
over flavor, color, and spinor indices is implied); $\tau_k$
($k=1,2,3$) are Pauli matrices; the baryon  chemical potential
$\mu_B$ in (\ref{1}) is responsible for the non-zero baryon density
of quark matter, whereas the isospin chemical potential $\mu_I$ is
switched on in order to study properties of quark matter at nonzero
isospin densities (in this case the densities of $u$ and $d$ quarks
are different). Evidently, the model (\ref{1}) is a generalization
of the two-dimensional Gross-Neveu model \cite{gn} with a single
massless quark color $N_c$-plet to the case of two quark flavors
and additional chemical potentials. As a result, we have in the case
under consideration a more complicated chiral symmetry group. Indeed,
at $\mu_I =0$ apart from the global color SU($N_c$) symmetry, the
Lagrangian (\ref{1}) is invariant under transformations from the
chiral $SU_L(2)\times SU_R(2)$ group. However, at $\mu_I \ne 0$ this
symmetry is reduced to $U_{I_3L}(1)\times U_{I_3R}(1)$, where
$I_3=\tau_3/2$ is the third component of the isospin operator
(here and above the subscripts $L,R$ mean that the corresponding
group acts only on the left, right handed spinors, respectively).
Evidently, this symmetry can also be presented
as $U_{I_3}(1)\times U_{AI_3}(1)$, where $U_{I_3}(1)$ is the isospin
subgroup and $ U_{AI_3}(1)$ is the axial isospin subgroup. Quarks
are transformed under these subgroups
as $q\to\exp (\mathrm{i}\alpha\tau_3) q$ and $q\to\exp (\mathrm{i}
\alpha\gamma^5\tau_3) q$, respectively. \footnote{\label{f1,1}
Recall for the following that~~
$\exp (\mathrm{i}\alpha\tau_3)=\cos\alpha
+\mathrm{i}\tau_3\sin\alpha$,~~~~
$\exp (\mathrm{i}\alpha\gamma^5\tau_3)=\cos\alpha
+\mathrm{i}\gamma^5\tau_3\sin\alpha$.}

The linearized version of the Lagrangian (\ref{1}), which contains
composite bosonic fields $\sigma (x)$ and $\pi_a (x)$ $(a=1,2,3)$,
has the following form (in what follows, we use the notation
$\mu\equiv\mu_B/3$ for the quark chemical potential):
\begin{eqnarray}
\tilde L\ds &=&\bar q\Big [\gamma^\nu\mathrm{i}\partial_\nu
+\mu\gamma^0
+ \frac{\mu_I}2\tau_3\gamma^0-\sigma
-\mathrm{i}\gamma^5\pi_a\tau_a\Big ]q
 -\frac{N_c}{4G}\Big [\sigma\sigma+\pi_a\pi_a\Big ].
\label{2}
\end{eqnarray}
From the Lagrangian (\ref{2}) one gets the equations 
for the bosonic fields 
\begin{eqnarray}
\sigma(x)=-2\frac G{N_c}(\bar qq);~~~\pi_a (x)=-2\frac G{N_c}(\bar q
\mathrm{i}\gamma^5\tau_a q).
\label{200}
\end{eqnarray}
Obviously, the Lagrangian (\ref{2}) is equivalent to the Lagrangian
(\ref{1}) when using the equations (\ref{200}).
Furthermore, it is clear from (\ref{200}) and footnote \ref{f1,1}
that the bosonic fields transform under the isospin $U_{I_3}(1)$ and
axial isospin $U_{AI_3}(1)$ subgroups in the following manner:
\begin{eqnarray}
U_{I_3}(1):&&\sigma\to\sigma;~~\pi_3\to\pi_3;~~\pi_1\to\cos
(2\alpha)\pi_1+\sin (2\alpha)\pi_2;~~\pi_2\to\cos (2\alpha)\pi_2-\sin
(2\alpha)\pi_1,\nonumber\\
U_{AI_3}(1):&&\pi_1\to\pi_1;~~\pi_2\to\pi_2;~~\sigma\to\cos
(2\alpha)\sigma+\sin (2\alpha)\pi_3;~~\pi_3\to\cos
(2\alpha)\pi_3-\sin (2\alpha)\sigma.
\label{201}
\end{eqnarray}
Starting from the theory (\ref{2}), one obtains in the leading order
of the large $N_c$-expansion (i.e. in the one-fermion loop
approximation) the following path integral expression for the
effective action ${\cal S}_{\rm {eff}}(\sigma,\pi_a)$ of the bosonic
$\sigma (x)$ and $\pi_a (x)$ fields:
$$
\exp(\mathrm{i}{\cal S}_{\rm {eff}}(\sigma,\pi_a))=
  N'\int[d\bar q][dq]\exp\Bigl(\mathrm{i}\int\tilde L\,d^2 x\Bigr),
$$
where
\begin{equation}
{\cal S}_{\rm {eff}}
(\sigma,\pi_a)
=-N_c\int d^2x\left [\frac{\sigma^2+\pi^2_a}{4G}
\right ]+\tilde {\cal S}_{\rm {eff}},
\label{3}
\end{equation}
$N'$ is a normalization constant.
The quark contribution to the effective action, i.e. the term
$\tilde {\cal S}_{\rm {eff}}$ in (\ref{3}), is given by:
\begin{equation}
\exp(\mathrm{i}\tilde {\cal S}_{\rm {eff}})=N'\int [d\bar
q][dq]\exp\Bigl(\mathrm{i}\int\Big [\bar q\mathrm{D}q\Big ]d^2
x\Bigr)=[\Det D]^{N_c}.
 \label{4}
\end{equation}
In (\ref{4}) we have used the notation $\mathrm{D}\equiv D\times
\mathrm{I}_c$, where $\mathrm{I}_c$ is the unit operator in the
$N_c$-dimensional color space and
\begin{equation}
D\equiv\gamma^\nu\mathrm{i}\partial_\nu +\mu\gamma^0
+ \frac{\mu_I}2\tau_3\gamma^0-\sigma -\mathrm{i}\gamma^5\pi_a\tau_a
\label{5}
\end{equation}
is the Dirac operator, which acts in the flavor-, spinor- as well as
coordinate spaces only. Using the general formula $\Det D=\exp {\rm
Tr}\ln D$, one obtains for the effective action the following
expression
\begin{equation}
{\cal S}_{\rm {eff}}(\sigma,\pi_a)
=-N_c\int
d^2x\left[\frac{\sigma^2+\pi^2_a}{4G}\right]-\mathrm{i}N_c{\rm
Tr}_{sfx}\ln D,
\label{6}
\end{equation}
where the Tr-operation stands for the trace in spinor- ($s$), flavor-
($f$) as well as two-dimensional coordinate- ($x$) spaces,
respectively. Using (\ref{6}), we obtain the thermodynamic
potential (TDP) $\Omega_{\mu,\mu_I}(\sigma,\pi_a)$
of the system:
\begin{eqnarray}
\Omega_{\mu,\mu_I}(\sigma,\pi_a)\!\!\!\!\!\!
&&\equiv -\frac{{\cal S}_{\rm {eff}}(\sigma,\pi_a)}{N_c\int
d^2x}~\bigg |_{~\sigma,\pi_a=\rm {const}}
=\frac{\sigma^2+\pi^2_a}{4G}+\mathrm{i}\frac{{\rm
Tr}_{sfx}\ln D}{\int d^2x}\nonumber\\
&&=\frac{\sigma^2+\pi^2_a}{4G}+\mathrm{i}{\rm
Tr}_{sf}\int\frac{d^2p}{(2\pi)^2}\ln\Big (\not\!p +\mu\gamma^0
+ \frac{\mu_I}2\tau_3\gamma^0-\sigma
-\mathrm{i}\gamma^5\pi_a\tau_a\Big ),
\label{7}
\end{eqnarray}
where the $\sigma$- and $\pi_a$ fields are now $x$-independent
quantities, and in the round brackets of (\ref{7}) just the momentum
space representation, $\bar D$, of the Dirac operator $D$ appears.
Evidently, ${\rm Tr}_{sf}\ln\bar D=$$\sum_i\ln\epsilon_i$, where the
summation over all four eigenvalues $\epsilon_i$ of the 4$\times$4
matrix $\bar D$ is implied and
\begin{eqnarray}
\epsilon_{1,2,3,4}=-\sigma\pm\sqrt{(p_0+\mu)^2-p_1^2-\pi_a^2+
(\mu_I/2)^2\pm \mu_I\sqrt{(p_0+\mu)^2-\pi_1^2-\pi_2^2}}.
\label{8}
\end{eqnarray}
Hence,
\begin{eqnarray}
\Omega_{\mu,\mu_I}(\sigma,\pi_a)&&\!\!\!\!\!\!
=\frac{\sigma^2+\pi^2_a}{4G}+\mathrm{i}\int\frac{d^2p}{(2\pi)^2}\ln
\Big (\epsilon_1\epsilon_2\epsilon_3\epsilon_4\Big )\nonumber\\
&&=\frac{\sigma^2+\pi^2_a}{4G}+\mathrm{i}\int\frac{d^2p}{(2\pi)^2}\ln
\Big\{\Big [(p_0+\mu)^2-\varepsilon_{+}^2\Big ]\Big
[(p_0+\mu)^2-\varepsilon_{-}^2\Big ]\Big\},
\label{9}
\end{eqnarray}
where
\begin{eqnarray}
\varepsilon_{\pm}=\sqrt{\Big (\sqrt{p_1^2+\sigma^2+\pi_3^2}
\pm\frac{\mu_I}2\Big )^2+\pi_1^2+\pi_2^2}.
\label{10}
\end{eqnarray}
The TDP $\Omega_{\mu,\mu_I}(\sigma,\pi_a)$ is symmetric under the
transformations $\mu\to -\mu$ and/or $\mu_I\to-\mu_I$.
Hence, it is sufficient to consider only the region $\mu\geq
0,\mu_I\geq 0$. In this case, one can integrate in (\ref{9})
over $p_0$ with the help of the formula
\begin{eqnarray}
\int\frac{dp_0}{2\pi}\ln\Big [(p_0+a)^2-b^2\Big ]= 
\frac{\mathrm{i}}2\Big\{|a-b|+|a+b|\Big\}
\label{1000}
\end{eqnarray}
(which is valid up to an infinite constant independent of
quantities $a$, $b$) and obtain:
\begin{eqnarray}
\Omega_{\mu,\mu_I}(\sigma,\pi_a)&&\!\!\!\!\!\!
=\frac{\sigma^2+\pi^2_a}{4G}-\int_{-\infty}^{\infty}\frac{dp_1}{4\pi}
\Big\{|\varepsilon_{+}-\mu|+|\varepsilon_{+}+\mu|+|\varepsilon_{-}-
\mu|+|\varepsilon_{-}+\mu|\Big\}\nonumber\\
&&=\frac{\sigma^2+\pi^2_a}{4G}-\int_{-\infty}^{\infty}\frac{dp_1}{
2\pi}\Big\{\varepsilon_{+}+\varepsilon_{-}+(\mu-\varepsilon_{+})
\theta (\mu-\varepsilon_{+})+(\mu-\varepsilon_{-})\theta
(\mu-\varepsilon_{-})\Big\}.
\label{11}
\end{eqnarray}
(To get the second line in (\ref{11}) we used the relations
$|\varepsilon_{\pm}+\mu|=\varepsilon_{\pm}+\mu$ and $\theta
(x)+\theta (-x)=1$.) In what follows we are going to investigate the
$\mu,\mu_I$-dependence of the global minimum point of the function
$\Omega_{\mu,\mu_I}(\sigma,\pi_a)$ vs $\sigma,\pi_a$. To simplify the
task, let us note that both the quasiparticle energies (\ref{10}) and
hence the TDP (\ref{11}) depend effectively only on the two
combinations $\sigma^2+\pi_3^2$ and
$\pi_1^2+\pi_2^2$ of the bosonic fields, which are invariants with
respect to the $U_{I_3}(1)\times U_{AI_3}(1)$ group, as is easily
seen from (\ref{201}). In this case, without loss of generality,
one can put $\pi_2=\pi_3=0$ in (\ref{11}),
and study the TDP as a function of only two variables,
$M\equiv\sigma$ and $\Delta\equiv\pi_1$.
Then the global minimum point of the 
TDP $\Omega_{\mu,\mu_I}(M,\Delta)$,
\begin{eqnarray}
\Omega_{\mu,\mu_I}(M,\Delta)&&\!\!\!\!\!\!
=\frac{M^2+\Delta^2}{4G}-\int_{-\infty}^{\infty}\frac{dp_1}{2\pi}\Big
\{E^+_{\Delta}+E^-_{\Delta}+(\mu-E^+_{\Delta})\theta
(\mu-E^+_{\Delta})+(\mu-E^-_{\Delta})\theta
(\mu-E^-_{\Delta})\Big\},
\label{12}
\end{eqnarray}
is the solution of the system of gap equations
\begin{eqnarray}
0=\frac{\partial\Omega_{\mu,\mu_I}(M,\Delta)}{\partial M}&\equiv&
\frac{M}{2G}-M\int_{-\infty}^{\infty}\frac{dp_1}{2\pi
E}\Big\{\frac{\theta(E_\Delta^+-\mu)E^+}{E_\Delta^+}+
\frac{\theta(E_\Delta^--\mu)E^-}{E_\Delta^-} \Big\},\nonumber\\
0=\frac{\partial\Omega_{\mu,\mu_I}
(M,\Delta)}{\partial\Delta}&\equiv&
\frac{\Delta}{2G}-\Delta\int_{-\infty}^{\infty}\frac{dp_1}{2\pi}\Big
\{\frac{\theta(E_\Delta^+-\mu)}{E_\Delta^+}+
\frac{\theta(E_\Delta^--\mu)}{E_\Delta^-} \Big\},
\label{13}
\end{eqnarray}
where $E_\Delta^\pm=\sqrt{(E^\pm)^2+\Delta^2}$,
$E^\pm=E\pm\frac{\mu_I}{2}$, and $E=\sqrt{p_1^2+M^2}$.
Evidently, the coordinates $M$ and $\Delta$ of the global minimum
point of the TDP (\ref{12}) supply us with two order parameters
(gaps), which are proportional to the ground state expectation values
of the form $\vev{\bar qq}$ and $\vev{\bar q\mathrm{i}\gamma^5\tau_1
q}$, respectively. If the gap $M$ is nonzero, then in the ground
state of the model the axial isospin symmetry $U_{AI_3}(1)$ (at
$\mu_I\ne 0$) is spontaneously broken down. Moreover, if the gap
$\Delta\ne 0$, then in the ground state, corresponding to the phase
with charged pion condensation, the isospin $U_{I_3}(1)$ symmetry is
spontaneously broken down.

\subsection{Pion condensation: the case of $\mu = 0$, $\mu_I\ne 0$}
\label{II.B}

Since at $\mu_I=0$, $\mu\ne 0$ the phase structure of different GN
models was reasonably well studied both in two dimensions
\cite{wolff,barducci,chodos,thies} and in three dimensions
\cite{zhou2} (in the last case the four-fermion theories are also
renormalizable), in this subsection we shall study for simplicity
the model (\ref{1}) only at zero quark chemical potential, i.e. at
$\mu = 0$, but $\mu_I\ne 0$. The corresponding TDP will be denoted as
$\Omega_{\mu_I}(M,\Delta)$ and can be obtained from (\ref{12}):
\begin{eqnarray}
\Omega_{\mu_I}(M,\Delta)
&=&\frac{M^2+\Delta^2}{4G}-\int_{-\infty}^{\infty}\frac{dp_1}{2\pi}
\Big\{E^+_{\Delta}+E^-_{\Delta}\Big\}\nonumber\\
&\equiv& V_0(\rho)-\int_{-\infty}^{\infty}\frac{dp_1}{2\pi}
\Big\{E^+_{\Delta}+E^-_{\Delta}-2\sqrt{\rho^2+p_1^2}\Big\},
\label{14}
\end{eqnarray}
where $\rho=\sqrt{M^2+\Delta^2}$ and $V_0(\rho)$ is the TDP of the
system in the vacuum, i.e. at $\mu_I =0$. In the vacuum the TDP is
usually called effective potential:
\begin{eqnarray}
V_0(\rho)&&\!\!\!\!\!\!
=\frac{\rho^2}{4G}-2\int_{-\infty}^{\infty}\frac{dp_1}{2\pi}
\sqrt{\rho^2+p_1^2}.
\label{15}
\end{eqnarray}
It is easily seen that both the TDP (\ref{14}) and the effective
potential (\ref{15}) are formally ultraviolet (UV) divergent
quantities. So, a few words are needed about the renormalization
procedure of the initial model. It is well known that all
four-fermion theories of the type (\ref{1}) are renormalizable in
two dimensional spacetime \cite{gn}. Moreover, in the leading order
of the large $N_c$ expansion only the coupling constant should be
renormalized in order to obtain finite (renormalized) expressions
for different quantities (see, e.g., \cite{kgk1}). It means that the
bare coupling constant $G$ of the model (\ref{1}) depends on the
cutoff parameter $\Lambda$, $G\equiv G(\Lambda)$, in such a way that
all UV divergences, arising from loop integrations when
$\Lambda\to\infty$, are compensated by corresponding terms of
$G(\Lambda)$. As a consequence, in the limit $\Lambda\to\infty$  one
must necessarily obtain finite expressions for physical quantities.
Of course, different renormalization procedures result in 
different expressions for the bare coupling constant $G(\Lambda)$.
However, physical consequences of the theory do not depend on the
concrete renormalization scheme. Taking this last remark into
account, let us next discuss how to obtain finite renormalized
expression for the TDP (\ref{14}). We see here two ways. On the one
hand, one could find an expression for the bare coupling constant $G$ 
such that the UV divergence, arising from the integral
in the first line of (\ref{14}), would be compensated by the term
with $G$. Evidently, in this case $G$ depends both on the cutoff
$\Lambda$ and $\mu_I$. However, we find it more convenient to
consider the second way. In this case, one should first of all note
that the integral in the second line of (\ref{14}) is a convergent
quantity, and the whole UV divergence is located in the effective
potential $V_0(\rho)$. Hence, there is a possibility to remove UV
divergences using a bare coupling constant which does not depend on
$\mu_I$. Namely, let us choose
\begin{eqnarray}
\frac{1}{2G}=\frac{2}{\pi}\int_0^\Lambda
dp_1\frac{1}{\sqrt{M_0^2+p_1^2}}=\frac{2}{\pi}\ln\left
(\frac{\Lambda+\sqrt{M_0^2+\Lambda^2}}{M_0}\right ),
\label{16}
\end{eqnarray}
where $M_0$ is the dynamical mass of quarks in the vacuum (for more
details, see Appendix \ref{ApA}). Then, substituting (\ref{16}) into
(\ref{15}) and restricting there the range of integration
by using the cutoff parameter $\Lambda$, it is possible to obtain
for $\Lambda>>M_0$ the expression (moreover, we omit an inessential
infinite constant independent of $\rho$):
\begin{eqnarray}
V_0(\rho)=
\frac{\rho^2}{2\pi}\left [\ln\left
(\frac{\rho^2}{M_0^2}\right )-1\right ].
\label{17}
\end{eqnarray}
Since $M_0$ might be considered as a free model parameter, it follows
from (\ref{16})-(\ref{17}) that the renormalization procedure of the
GN model is accompanied by the dimensional transmutation phenomenon.
Indeed, in the initial unrenormalized expressions both for
$\Omega_{\mu_I}(M,\Delta)$ and $V_0(\rho)$ (see (\ref{14}) and
(\ref{15}), respectively) the dimensionless coupling constant $G$ is
present, whereas after renormalization the effective potential
(\ref{17}) is characterized by a dimensional free model parameter
$M_0$.

Due to the relation (\ref{17}), one can
show that the gap equations for the renormalized TDP (\ref{14})
might have no more than three different solutions. Two of them,
$(M=0,\Delta=0)$ and $(M=0,\Delta=M_0)$, are present at arbitrary
values of $\mu_I\ge 0$, whereas the third one, $(M=M_0,\Delta=0)$,
appears only at $\mu_I<M_0\sqrt{2}$. However, for arbitrary $\mu_I
>0$ a global minimum point of the TDP $\Omega_{\mu_I}(M,\Delta)$
lies at the point $(M=0,\Delta =M_0)$. This means that in the model
(\ref{1}) the isospin symmetry is always broken down and a charged
pion condensate which is equal to the quark mass $M_0$ in the vacuum,
is created if $\mu_I >0$.

Since in the vacuum case ($\mu=0, \mu_I=0$) chiral symmetry is
spontaneously broken down in the model (\ref{1}), there must exist
three massless Nambu--Goldstone bosons which are pions, i.e.
$m_{\pi}=0$. So, we have proved that in the framework of the model
(\ref{1}) the pion condensation phase is realized at $\mu_I>m_{\pi}$,
where $m_{\pi}$ is the pion mass in the vacuum. Just the same phase
structure is predicted by QCD at $\mu=0$, $\mu_I\ne 0$ \cite{son}. In
contrast, in the framework of (3+1)-dimensional NJL-type models the
pion condensation is not allowed for sufficiently high values of the
isospin chemical potential \cite{frank,ek,andersen}. This fact
supports the statement made in the Introduction that the NJL approach
is only valid at rather small energies (chemical potentials).
Moreover, we have once more demonstrated that in the leading order of
the large $N_c$-expansion the two-dimensional GN models are a quite
good theoretical laboratory for qualitative QCD investigations. So we
are in a position to believe that the results obtained in the next
sections are also inherent to QCD.

\section{The case of  $R^1\times S^1$ spacetime and $\mu_I\ne 0$}
\label{III}

In the present section we continue the investigation of the charged
pion condensation, this time under the influence of the finite
volume occupied by the system. This is obviously a reasonable task,
since all physical effects take place in restricted space regions.
The consideration of the problem is significantly simplified in the
framework of the two-dimensional model (\ref{1}) at $\mu_I\ne 0$,
which is again justified by its similarity to QCD. So we put a system
with Lagrangian (\ref{1}) into a restricted space region of the form
$0\le x\le L$ (here $x$ is the space coordinate). It is well known
that in this case the consideration is equivalent to the
investigation of the model in a spacetime with nontrivial topology
$R^1\times S^1$ and with quantum fields, satisfying some boundary
conditions of the form
\begin{eqnarray}
q(t,x+L)=e^{i\pi\alpha}q(t,x),
\label{18}
\end{eqnarray}
where $0\le\alpha < 2$, $L$ is the length of the circumference $S^1$,
and the variable $x$ means the path along it. Below, we shall use
only two values of the parameter $\alpha$: $\alpha=0$ for periodic
boundary conditions and $\alpha=1$ for the antiperiodic one.

As a consequence, to obtain the thermodynamic potential
$\Omega_{L\mu_I}(M,\Delta)$ of the initial system placed in the
restricted domain $0\le x\le L$ and at $\mu_I\ne 0$, one must 
simply replace the integration in (\ref{14}) by an infinite series,
using the rule:
\begin{eqnarray}
\int_{-\infty}^{\infty}\frac{dp_1}{2\pi}f(p_{1})\to\frac
1L\sum_{n=-\infty}^{\infty}f(p_{1n}),~~~~p_{1n}=\frac{\pi}{L}(2n+
\alpha),~~~n=0,\pm 1, \pm 2,...
\label{19}
\end{eqnarray}
Moreover, instead of $V_0(\rho)$ it is necessary to use the effective
potential $V_{L}(\rho)$ of the model in the vacuum (see Appendix
\ref{ApA}). As a result, the TDP (\ref{14}) will be replaced by the
corresponding expression for the spacetime of the form $R^1\times
S^1$, i.e. 
\begin{eqnarray}
\Omega_{L\mu_I}(M,\Delta)
&=&V_L(\rho)-\frac 1L\sum_{n=-\infty}^{\infty}
\left\{\sqrt{\left
(\sqrt{M^2+\frac{\pi^2}{L^2}(2n+\alpha)^2}+\frac{\mu_I}{2}
\right )^2+\Delta^2}\right.\nonumber\\
&+&\left.\sqrt{\left
(\sqrt{M^2+\frac{\pi^2}{L^2}(2n+\alpha)^2}-\frac{\mu_I}{2}
\right )^2+\Delta^2}-2\sqrt{\rho^2+\frac{\pi^2}{L^2}(2n+\alpha)^2}
\right\},
\label{20}
\end{eqnarray}
where $\rho=\sqrt{M^2+\Delta^2}$, and the function $V_L(\rho)$ is
defined in (\ref{A9}). In what follows, it will be convenient to
use the dimensionless quantities
\begin{eqnarray}
\lambda=\frac{\pi}{LM_0},~~\nu=\frac{\mu_I}{2M_0},~~m=\frac{M}{M_0},~
~\delta=\frac{\Delta}{M_0},~~{\mathfrak
O}_{\lambda\nu}(m,\delta)=\frac{\pi}{M_0^2}\Omega_{L\mu_I}(M,
\Delta),
\label{21}
\end{eqnarray}
where $M_0$ is the dynamical quark mass in the vacuum. Moreover,
since the phase structure of the model in the two particular cases
$L=\infty$, $\mu_I\ne 0$  and $L\ne\infty$, $\mu_I=0$ was already
considered in section \ref{II.B} and in the Appendix
\ref{ApA}, we will now investigate the phase structure only at
$\lambda>0,\nu>0$.

\begin{figure}
   \noindent
 \centering
 $
 \begin{array}{cc}
 \epsfig{file=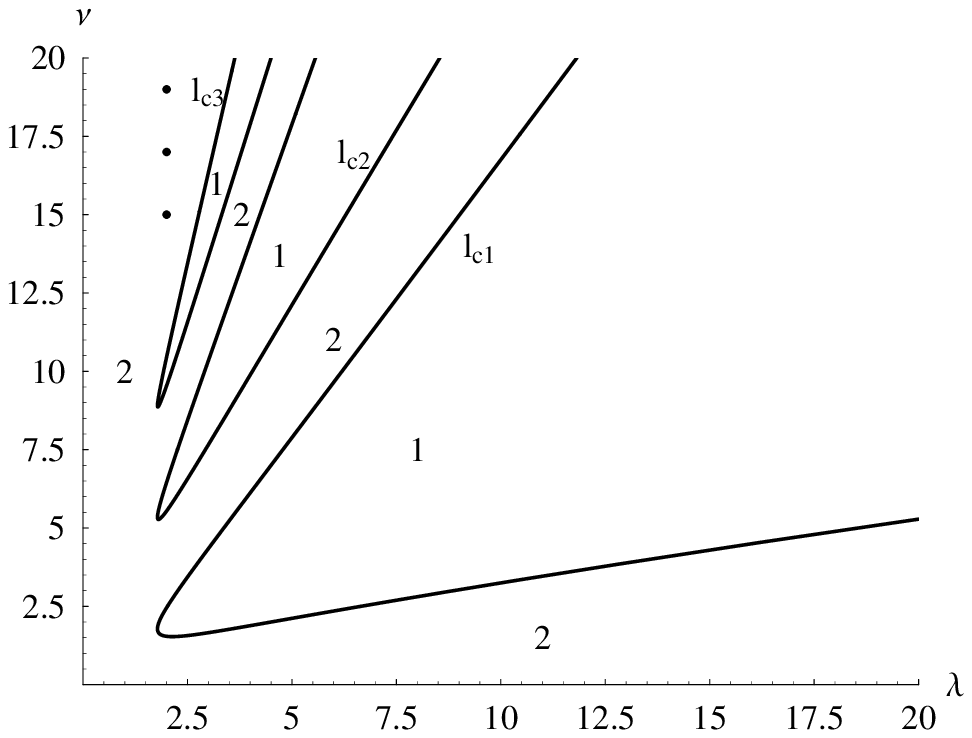,width=8cm}~~~&~~~
 \epsfig{file=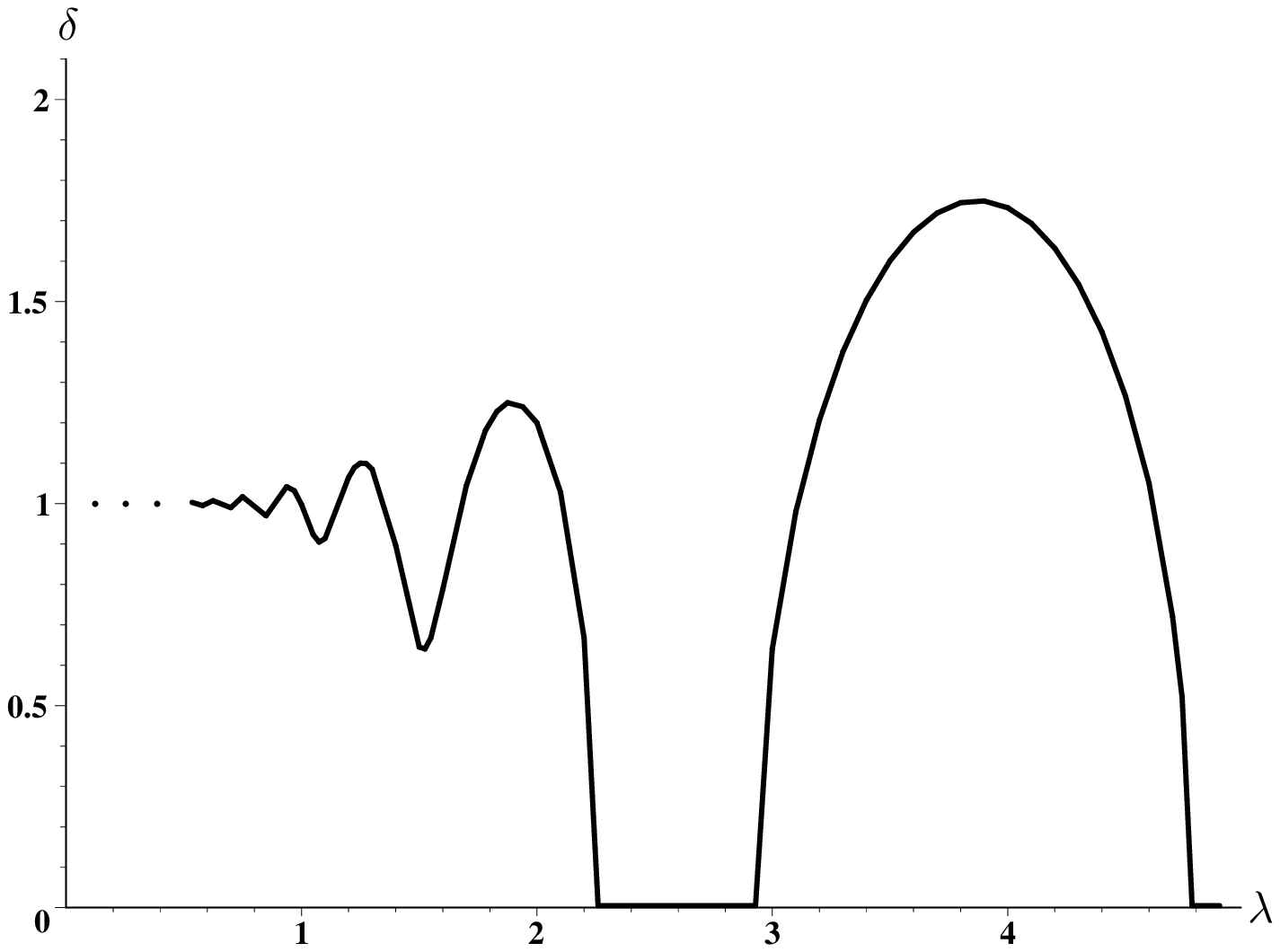,width=7.5cm}
 \end{array}
 $
 \caption{ {\bf The periodic case}: In the left picture
 the phase portrait of the model is represented in terms of
 $(\lambda,\nu)$, where number 1 denotes the symmetric phase, number
 2 -- the pion condensed phase. In the right picture the behavior of
 the gap $\delta$ vs $\lambda$ is depicted at $\nu=7.5$.}
   \label{pp}
\end{figure}

\subsection{The case of periodic boundary conditions}
\label{III.A}

In this case $\alpha =0$, and in terms of the dimensionless
quantities (\ref{21}) the TDP (\ref{20}) can be rewritten in the
following explicit form
\begin{eqnarray}
{\mathfrak
O}_{\lambda\nu}(m,\delta)&=&(m^2+\delta^2)[\ln(4\lambda)-\gamma]-
\lambda\sqrt{(m+\nu)^2+\delta^2}-\lambda\sqrt{(m-\nu)^2+\delta^2}
\nonumber\\&-&2\lambda\sum_{n=1}^{\infty}\left\{\sqrt{\left
(\sqrt{m^2+(2n\lambda)^2}+\nu\right )^2+\delta^2}+\sqrt{\left
(\sqrt{m^2+(2n\lambda)^2}-\nu\right )^2+\delta^2}-4n\lambda-
\frac{m^2+\delta^2}{2n\lambda}\right\},
\label{22}
\end{eqnarray}
where $\gamma=0.577...$ is the Euler's constant \cite{poisson}. We
consider the TDP (\ref{22}) as a function of two variables,
$m,\delta$. Moreover, $\nu,\lambda$ are free parameters there. Since
the information about the phase structure of the model in the case
of the periodic boundary conditions is contained in the global
minimum point of the function (\ref{22}) vs $m,\delta$,
it is first of all necessary to study the gap equations 
and then to investigate the behavior of the global minimum point vs
parameters $\nu,\lambda$. In particular, it is possible to show that
for each fixed point of the plane $(\lambda,\nu)$ (with $\nu > 0$
and $\lambda\ge 0$) the global minimum point of the TDP (\ref{22})
might be located at two different points only, i) $(m=0,\delta=0)$
and ii) $(m=0,\delta\ne 0)$, where the nonzero gap $\delta$ is the
solution of the equation $\partial{\mathfrak
O}_{\lambda\nu}(0,\delta)/ \partial (\delta^2)=0$. The point i)
corresponds to the $U_{I_3}(1)\times U_{AI_3}(1)$ symmetric phase of
the model (without charged pion condensation).
On the other hand, if the global minimum of the function (\ref{22})
is situated at the point ii), then in the ground state of the model
the isospin symmetry $U_{I_3}(1)$ is spontaneously broken down, and
the pion condensation takes place.
Let us denote by $l_c$ the critical curve which separates
the region of the $(\lambda,\nu)$ plane with symmetric phase from
the points $(\lambda,\nu)$, corresponding to the pion condensed
phase of the model. Since in each point of the curve $l_c$ there is
a phase transition of the second order from the symmetric phase to
the pion condensed one and vice versa, the gap $\delta$ must vanish
on this curve. So the critical curve $l_c$ is defined by the
following equation
\begin{eqnarray}
l_{c}:~~~\frac{\partial{\mathfrak
O}_{\lambda\nu}(m,\delta)}{\partial (\delta^2)}\Big
|_{m,\delta=0}\equiv\ln(4\lambda)-\gamma-\frac\lambda\nu
-\sum_{n=1}^{\infty}\left
\{\frac{\lambda}{2n\lambda+\nu}+\frac{\lambda}{|2n\lambda-\nu|}-
\frac{1}{n}\right\}=0.
\label{23}
\end{eqnarray}
To represent the curve $l_c$ in the plane $(\lambda,\nu)$, it is
convenient to divide this plane into an infinite set of regions
$\omega_k$:
\begin{eqnarray}
(\lambda,\nu)=
\bigcup\limits^{\infty}_{k=1}\omega_k;~~~~~
\omega_k=\{(\lambda,\nu):2\lambda (k-1) \le \nu\le 2\lambda k\}.
\label{24}
\end{eqnarray}
In accordance with the division (\ref{24}), the critical curve $l_c$
can also be presented as a set of pieces,
$l_c=\bigcup\limits^{\infty}_{k=1}l_{ck}$.
Obviously, each piece $l_{ck}$ of the whole critical curve $l_c$
lies inside the corresponding $k$-th region $\omega_k$ and obeys the
following equation ($k>1$)
\begin{eqnarray}
l_{ck}:~~~\ln(4\lambda)-\gamma-\frac\lambda\nu-
\sum_{n=1}^{k-1}\left
\{\frac{\lambda}{2n\lambda+\nu}+\frac{\lambda}{\nu-2n\lambda}-
\frac{1}{n}\right\}-\sum_{n=k}^{\infty}\left
\{\frac{\lambda}{2n\lambda+\nu}+\frac{\lambda}{2n\lambda-\nu}-
\frac{1}{n}\right\}=0.
\label{25}
\end{eqnarray}
For $k=1$ the part $l_{c1}$ obeys the equation (\ref{23}) with
omitted absolute value symbols. Performing the summations in
(\ref{23}) or (\ref{25}), one can find for each piece $l_{ck}$ of the
critical curve $l_c$ the following equation ($k\ge 1$)
\begin{eqnarray}
\label{26} l_{ck}:&&~~~
2\ln(4\lambda)+2\psi\left
(k-\frac{\nu}{2\lambda}\right )-\psi\left
(1-\frac{\nu}{2\lambda}\right )+\psi\left (\frac{\nu}{2\lambda}\right
)=0,\end{eqnarray}
which is valid only at $2\lambda (k-1)\le\nu\le 2\lambda k$. Here
$\psi(x)$ is the logarithmic derivative of the Euler's $\Gamma
(x)$ function \cite{poisson}.
Before drawing the critical curve $l_c$, we would like to point out
one its peculiarity. Using the well-known property of the $\psi(x)$
function, $\pi\cot (\pi x)=\psi(1-x)-\psi(x)$, as well as the
periodicity of $\cot (\pi x)$, the equation (\ref{26}) can be reduced
to the following one:
\begin{eqnarray}
\label{27} l_{ck}:&&~~~
2\ln(4\lambda)=-2\psi (z)-\pi\cot (\pi z)\equiv F(z), \end{eqnarray}
where $z=k-\nu/(2\lambda)$ and $0\le z\le 1$. Since the absolute
minimum of the function $F(z)$ from (\ref{27})
corresponds to the point $z=1/2$, each branch $l_{ck}$ of the
critical curve lies to the right of the vertical line
$\lambda=\lambda_c$ (in the plane $(\lambda,\nu)$), where
$2\ln(4\lambda_c)=F(1/2)$, i.e. $\lambda_c=e^\gamma\approx 1.78$. All
the branches of the critical curve $l_c$ as well as the phase
portrait of the initial model in terms of $(\lambda,\nu)$ are
presented in Fig. 1 (left picture). Clearly, there is a strip $0\le
\lambda <\lambda_c$ which lies, as a whole, inside the region,
corresponding to the pion condensed phase.

In the right picture of Fig. 1 the behavior of the pion condensation
gap $\delta$ vs $\lambda$ is depicted at $\nu=7.5$. It is easily
seen that this quantity oscillates as a function of $\lambda$.
However, the amplitude of this oscillations is a rapidly decreasing
function of $\lambda$ when $\lambda\to 0$. Similar oscillations of
different physical quantities such as gaps, critical curves, particle
densities etc vs $\lambda$ were also observed in some NJL-type models
with one compactified space coordinate, but in a qualitatively
alternative case with nonzero baryonic chemical potential
\cite{vshivtsev1}. Moreover, oscillating phenomena as functions of
curvature are inherent to NJL models in the Einstein universe, i.e.
in the curved spacetime of the form $R^1\times S^3$ \cite{tyukov}. In
Fig. 2 the behavior of the gap $\delta$ vs $\nu$ is depicted at
$\lambda =1$ (left picture) and $\lambda =1.7$ (right picture).
Concerning this type of oscillations of the gap $\delta$, it is
necessary to note first of all that its period is equal to
$2\lambda$. Moreover, it is clear from Fig. 2, and this fact is
supported by numerical calculations, that the amplitude of the
oscillations of the quantity $\delta$ vs $\nu$ is a very slowly
decreasing function of $\nu$. Finally, it is evident that the smaller
$\lambda$, the smaller the amplitude of this $\nu$-oscillations of
the gap $\delta$.

\begin{figure}
   \noindent
 \centering
 $
 \begin{array}{cc}
 \epsfig{file=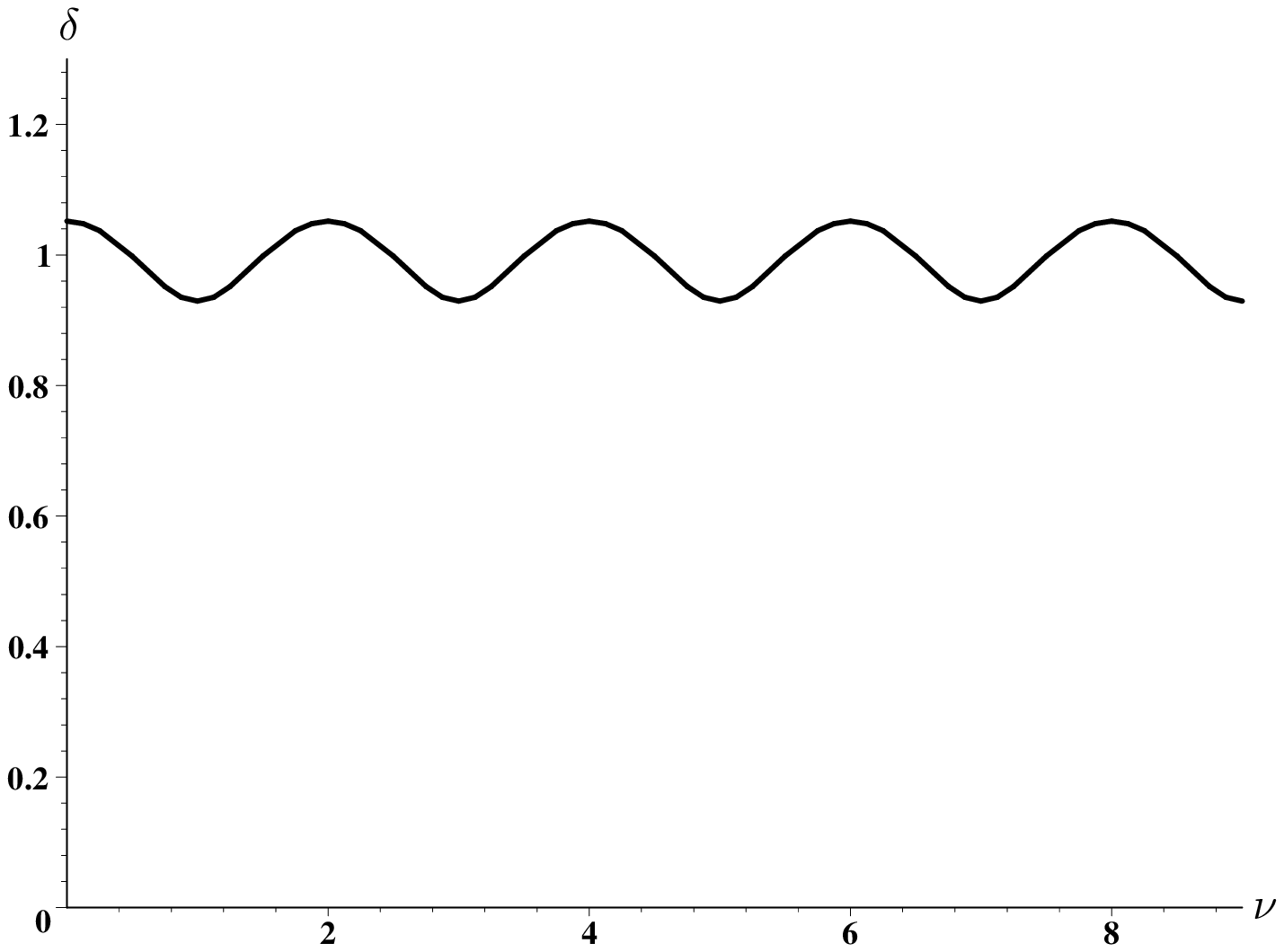,width=7.5cm}~~~&~~~
 \epsfig{file=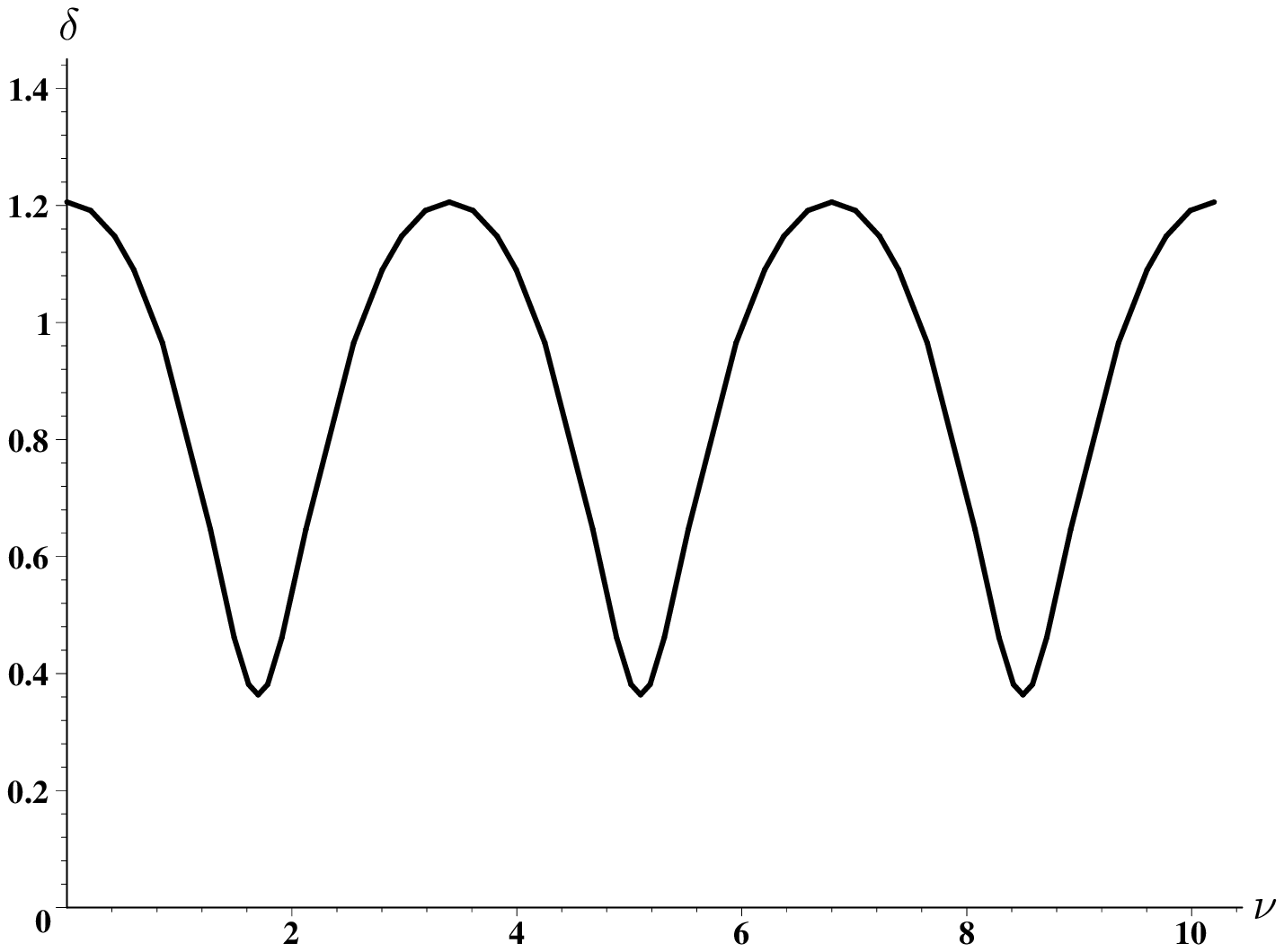,width=7.5cm}
 \end{array}
 $
 \caption{{\bf The periodic case}: The behavior of the gap $\delta$
 vs $\nu$ at $\lambda =1$ (left picture) and $\lambda =1.7$ (right
 picture).}
\end{figure}

\subsection{The case of antiperiodic boundary conditions}
\label{III.B}

In this case $\alpha=1$, so in (\ref{20}) instead of $V_L$ one should
use the effective potential (\ref{A11}). Then in terms of the
quantities (\ref{21}) we have
\begin{eqnarray}
{\mathfrak
O}_{\lambda\nu}(m,\delta)&=&(m^2+\delta^2)[\ln(\lambda)-\gamma]-
2\lambda\sum_{n=0}^{\infty}\left\{\sqrt{\left
(\sqrt{m^2+(2n+1)^2\lambda^2}+\nu\right )^2+\delta^2}
\right.\nonumber\\&&+\left.\sqrt{\left
(\sqrt{m^2+(2n+1)^2\lambda^2}-\nu\right )^2+\delta^2}-2(2n+1)\lambda-
\frac{m^2+\delta^2}{(2n+1)\lambda}\right\}.
\label{30}
\end{eqnarray}
The critical curve $l_c$ which divides the parameter plane
$(\lambda,\nu)$ into a region with symmetric phase and the region,
corresponding to a pion condensed phase, is now defined by the
following equation
\begin{eqnarray}
l_{c}:~~~\frac{\partial{\mathfrak
O}_{\lambda\nu}(m,\delta)}{\partial (\delta^2)}\Big
|_{m,\delta=0}\equiv\ln(\lambda)-\gamma
-\sum_{n=0}^{\infty}\left
\{\frac{\lambda}{(2n+1)\lambda+\nu}+\frac{\lambda}{|(2n+1)\lambda-\nu
|}-\frac{2}{2n+1}\right\}=0.
\label{31}
\end{eqnarray}
As in the case with periodic boundary conditions, for solving
the equation (\ref{31}) it is convenient to represent the parameter
$(\lambda,\nu)$-plane as the union of $\omega_k$ regions,
$(\lambda,\nu)= \bigcup\limits^{\infty}_{k=0}\omega_k$, where
\begin{eqnarray}
\omega_0=\{(\lambda,\nu):0 \le \nu\le \lambda \},~~~
\omega_k=\{(\lambda,\nu):(2k-1)\lambda \le \nu\le (2k+1)\lambda
\}~~~~\mbox{for} ~~~k\ge 1.
\label{32}
\end{eqnarray}
Accordingly, in this case the critical curve $l_c$ is composed of
different pieces, i.e. $l_c=\bigcup\limits^{\infty}_{k=0}l_{ck}$,
where $l_{ck}$ is the part of $l_c$, arranged in the corresponding
region $\omega_k$. Obviously, the equation for $l_{c0}$ is just the
equation (\ref{31}) with omitted absolute value symbols. However, the
equations for $l_{ck}$ ($k\ge 1$) look like
\begin{eqnarray}
l_{ck}:&&~~~\ln(\lambda)-\gamma-\sum_{n=1}^{k-1}\left
\{\frac{\lambda}{(2n+1)\lambda+\nu}+\frac{\lambda}{\nu-(2n+1)
\lambda}-\frac{2}{2n+1}\right\}\nonumber\\
&&~~~~~~~~~~~~~~~~~~~~~~~~~~~~~~~~~~~~~-\sum_{n=k}^{\infty}\left
\{\frac{\lambda}{(2n+1)\lambda+\nu}+\frac{\lambda}{(2n+1)\lambda-\nu}
-\frac{2}{(2n+1)}\right\}=0.
\label{33}
\end{eqnarray}
Summing in (\ref{31}) or (\ref{33}) with the help of a program of
analytical calculations, it is possible to obtain the more
concise form of the equations for different pieces $l_{ck}$ ($k\ge
0$) of the critical curve:
\begin{eqnarray}
\label{34} l_{ck}:&&~~~ 2\ln(4\lambda)+2\psi\left (k+\frac
12-\frac{\nu}{2\lambda}\right )-\psi\left (\frac
12-\frac{\nu}{2\lambda}\right )+\psi\left (\frac
12+\frac{\nu}{2\lambda}\right )=0.\end{eqnarray}
As in section \ref{III.A}, the equation (\ref{34}) can be
transformed to a formally $\omega_k$-independent expression
\begin{eqnarray}
\label{35} l_{ck}:&&~~~ 2\ln(4\lambda)=-2\psi (z)-\pi\cot (\pi z),
\end{eqnarray} where $z=k+\frac 12-\nu/(2\lambda)$ and $0\le z\le
\frac 12$ for $k=0$, whereas $0\le z\le 1$ for $k\ge 1$. Note
that the equation (\ref{35}) coincides with (\ref{27}) except for the
different $\nu,\lambda$ dependence of the variable $z$. In Fig. 3
(left picture) the first several branches $l_{ck}$ of the whole
critical curve $l_c$, which divides the $(\lambda,\nu)$ plane into a
region with pion condensed phase (the number 2 in the figure) and a
region corresponding to a symmetric phase (the number 1 in the
figure), are represented. Note that the strip
$0\le\lambda\le\lambda_c$ of the plane belongs to the region 2 with
the pion condensation phase.

In the right picture of Fig. 3 as well as in Fig. 4 the oscillating
behavior of the pion condensation gap $\delta$ vs $\lambda$ and,
correspondingly, vs $\nu$ is depicted. The properties of these
oscillations are the same as in the periodic case. Namely, at fixed
$\nu$ the gap $\delta$ is a quickly damping oscillating function of
$\lambda$ when $\lambda\to 0$, whereas at fixed $\lambda$ the gap
$\delta$ oscillates at $\nu\to\infty$ with a very weak damping.

\begin{figure}
   \noindent
 \centering
 $
 \begin{array}{cc}
 \epsfig{file=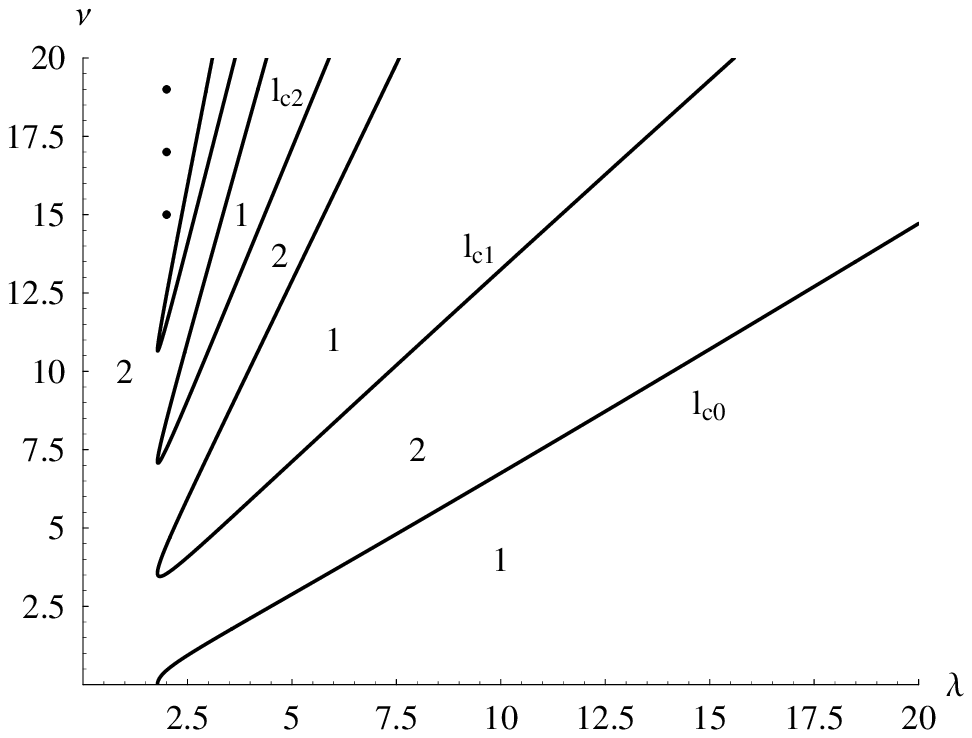,width=8cm}~~~&~~~
 \epsfig{file=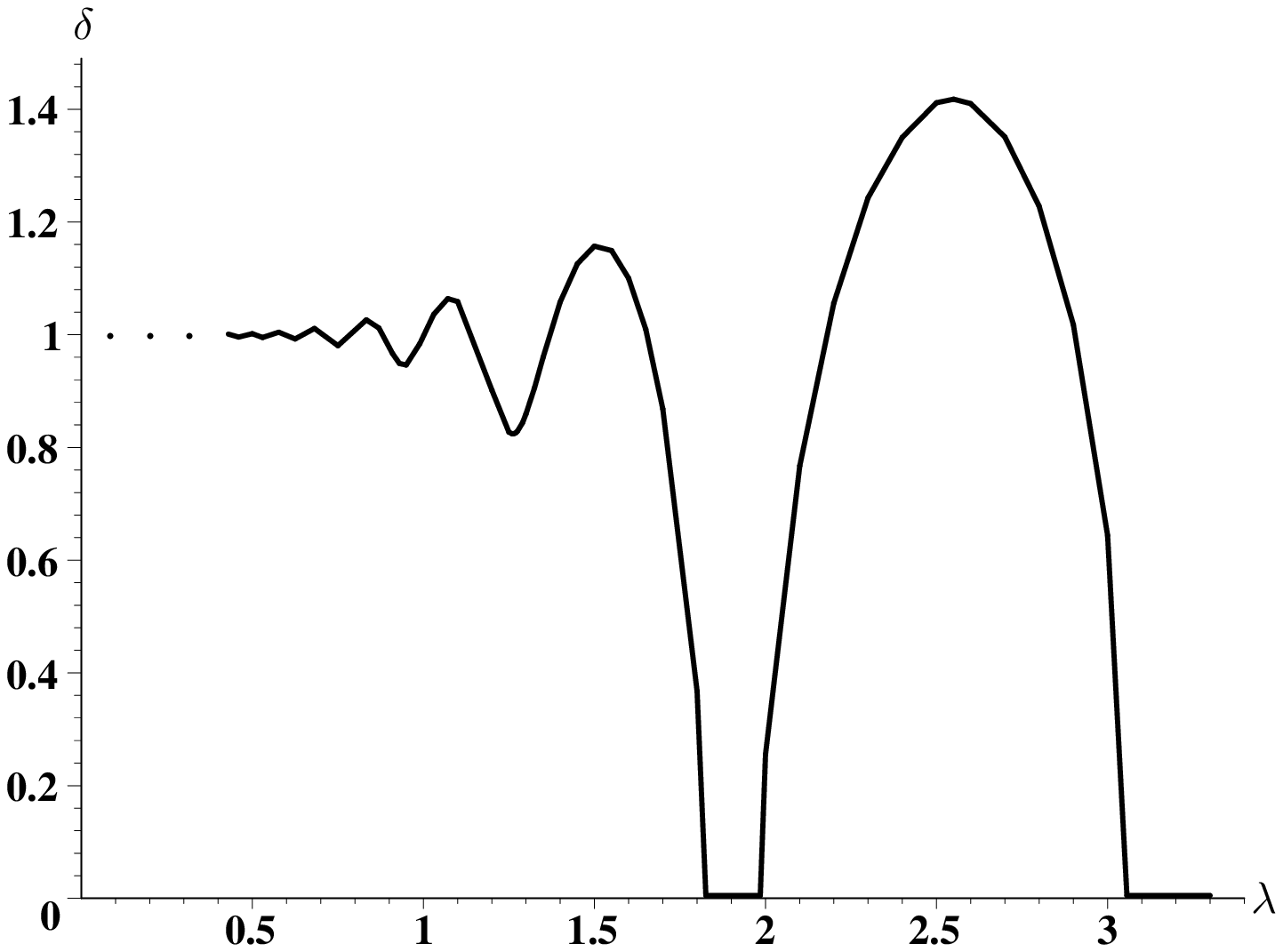,width=7.5cm}
 \end{array}
 $
 \caption{{\bf The antiperiodic case}: The phase portrait
 (left picture) and the gap $\delta$ at $\nu=7.5$ (right picture).
 The notations are the same as in Fig. 1.}
\end{figure}

\begin{figure}[ht]
   \noindent
 \centering
 $
 \begin{array}{cc}
 \epsfig{file=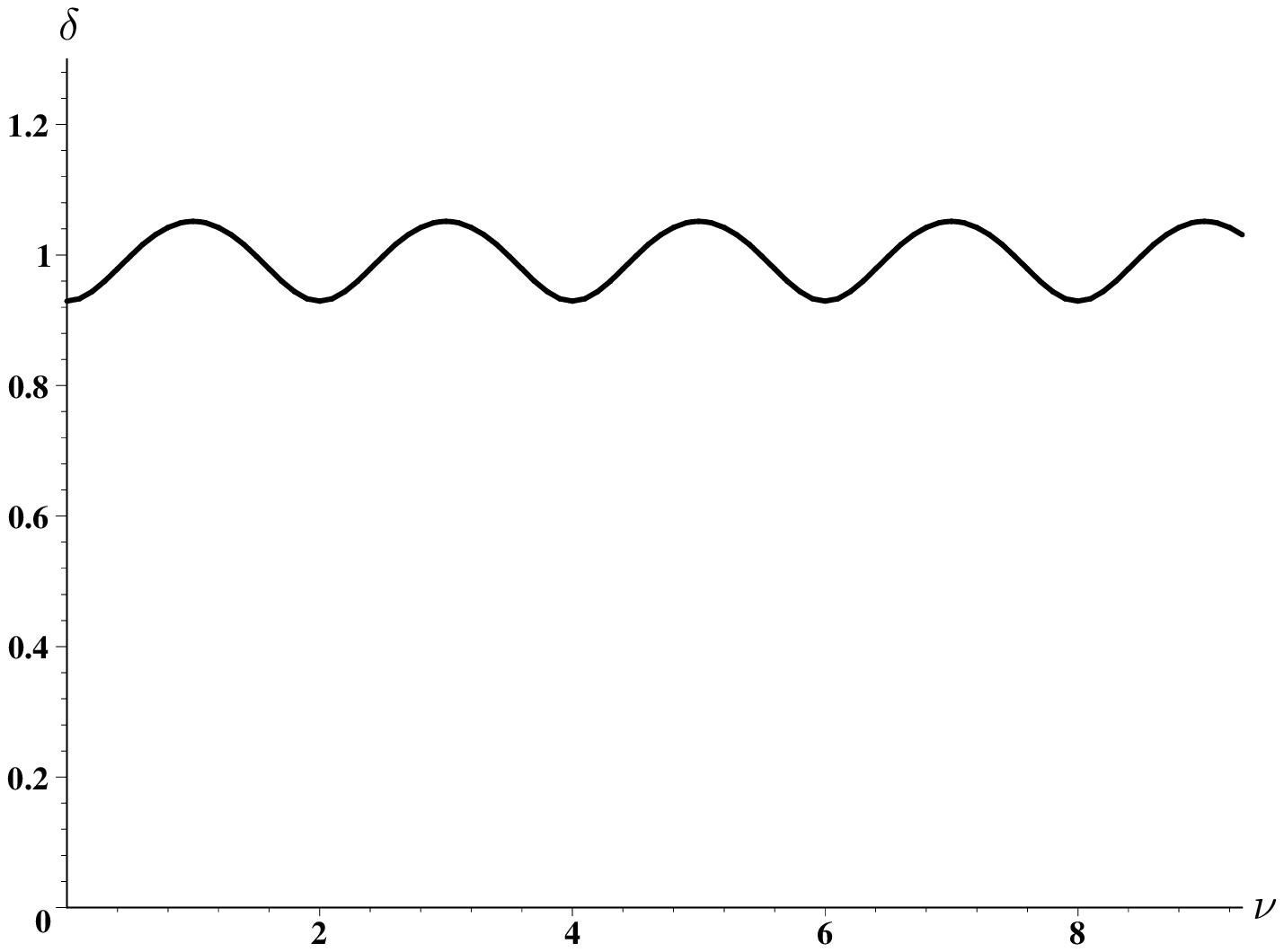,width=7.5cm}~~~&~~~
 \epsfig{file=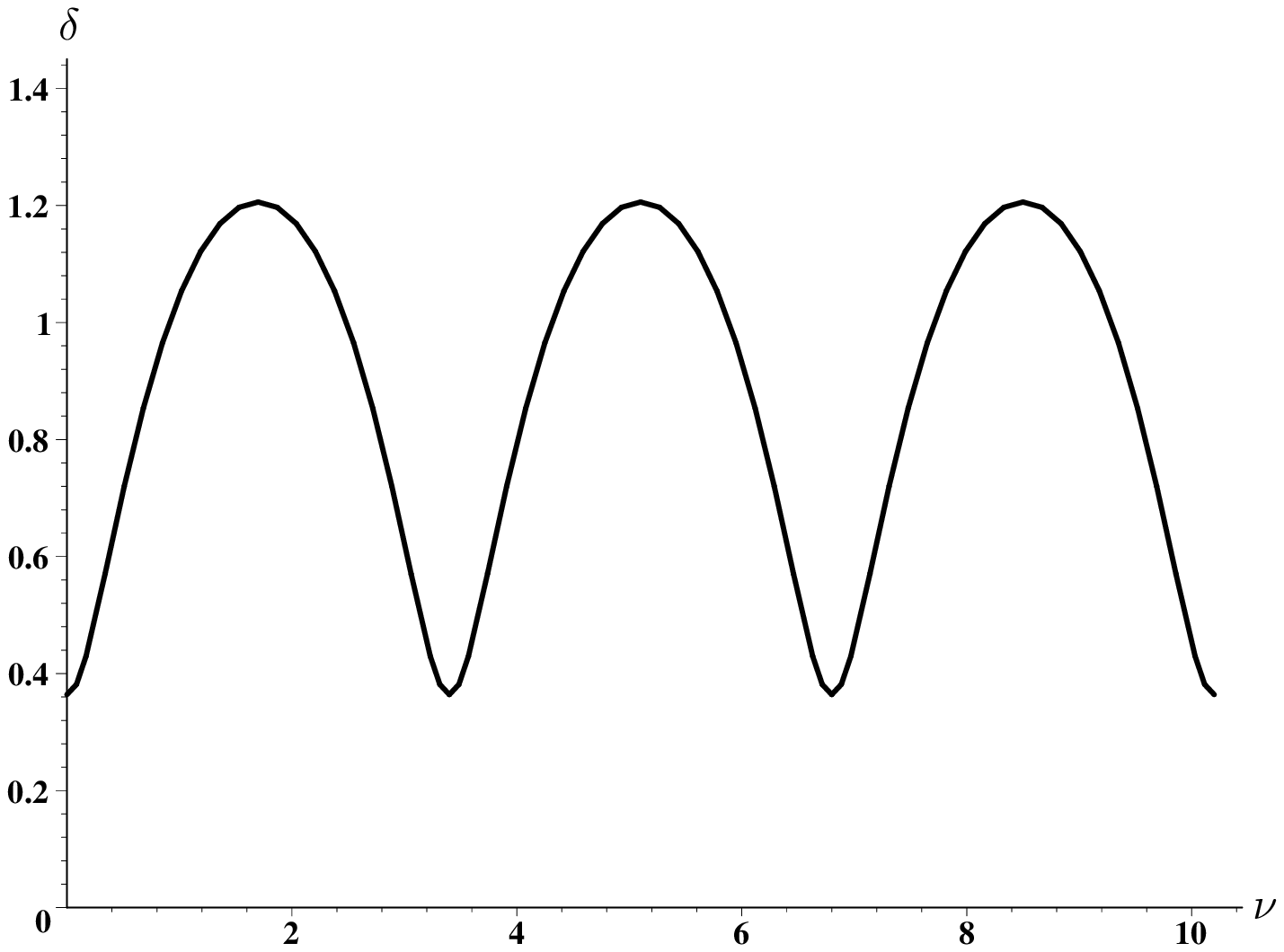,width=7.5cm}
 \end{array}
 $
 \caption{{\bf The antiperiodic case}: The behavior of the gap
 $\delta$ vs $\nu$ at $\lambda =1$ (left picture) and $\lambda =1.7$
 (right picture).}
\end{figure}

\section{Summary and discussion}

In the present paper we have studied the phase structure of a
two-dimensional GN model at nonzero isospin chemical potential
$\mu_I$ and in the spacetime $R^1\times S^1$ with nontrivial
topology, when the space coordinate is compactified into a
circumference of a finite length $L$. The consideration is performed
in the leading order of the large-$N_c$ expansion technique.
\footnote{
It should be noted that the problem of the IR-behavior of the
correlation function of quantum fluctuations in two-dimensional QFT
models was extensively discussed in literature with relation to the
Coleman-Mermin-Wagner theorem. One may mention the papers on the
2-dimensional GN model \cite{VSMT,barducci,chodos,thies},
where this problem has been investigated and it was demonstrated that
for the limit of infinite $N_c$ this theorem is not valid, and hence
spontaneous symmetry breaking may take place. }

It turns out that in the case with $L=\infty$ the pion condensed
phase is realized in the model at arbitrary nonzero values of
$\mu_I$. In this phase the corresponding order parameter, the pion
condensate $\Delta$, does not depend on the isospin chemical
potential $\mu_I$ and is equal to $M_0$, i.e. to the dynamical quark
mass in the vacuum. The same phase structure at $\mu_I\ne 0$
occurs in the chirally symmetric QCD, where pions are massless
particles, so one more common property is found which is shared both
by the GN model and QCD. As a result, the assurance that the
finite size ($L\ne\infty$) effects of the GN model are also inherent
to compactified QCD at $\mu_I\ne 0$ is raised.

If $L$ is finite, then the phase portraits of the model in terms of
$\lambda\sim 1/L$ and $\nu\sim\mu_I$ are found for the case of
periodic (see Fig.1) and antiperiodic (see Fig. 3) boundary
conditions. Among the most interesting properties of these phase
diagrams is the fact that the strip $0\le\lambda<\lambda_c\approx
1.78$ lies as a whole inside the pion condensed phase.
We have shown also that the pion condensed gap $\delta$ is an
oscillating function vs both $\lambda$ (at fixed $\nu$) and $\nu$ (at
fixed $\lambda$). The same is true for other thermodynamic quantities
of the model such as pressure, particle densities etc, and
is inherent also to the (3+1)-dimensional NJL models with curved
spacetimes \cite{tyukov} or spacetimes with non-trivial topology
\cite{vshivtsev1}.

One more interesting issue should also be mentioned.
Although in the present paper the spatially uniform pion
condensation was assumed for simplicity, however, for a sufficiently
high values of $\mu_I$, the pion superfluidity with inhomogeneous
condensate might be realized in isotopically asymmetric and spatially
infinite quark matter systems \cite{jin}.  A detailed investigation
of this possibility in the case of finite space volume is outside the
scope of this paper and should be left for further studies in the
framework of different QCD-like models including the GN model.

\section*{Acknowledgments}
Two of the authors (V.Ch.Zh. and A.V.T.) 
are grateful to Prof. M.~Mueller-Preussker for 
his kind hospitality during their stay in the particle theory
group at the Institute of Physics of Humboldt-University, where 
part of this work has been done, and also to DAAD for financial
support. This work was also supported in part by DFG-grant 436 RUS
113/477.

\appendix

\section{Effective potential in the vacuum ($\mu_I=0$)}
\label{ApA}

Note that in the vacuum and in the $R^1\times R^1$ spacetime, the
expression $V_0$ for the effective potential of the initial model
(\ref{1}) can be found starting from the TDP (\ref{9}) at
$\mu=\mu_I=0$, where without loss of generality it is possible to put
$\pi_a=0$ ($a=1,2,3$):
\begin{eqnarray}
V_0(\sigma)
=\frac{\sigma^2}{4G}+i\int\frac{d^2p}{(2\pi)^2}\ln \Big [(
p_0^2-p_1^2-\sigma^2)^2\Big ]
=\frac{\sigma^2}{4G}-2\int_{-\infty}^{\infty}\frac{dp_1}{2\pi}
\sqrt{\sigma^2+p_1^2}.
\label{A1}
\end{eqnarray}
Here the second equality is obtained with the help of formula
(\ref{1000}). Since in (\ref{A1}) the last integral is an UV
divergent one, we regularize it by cutting off the integration
region, i.e. supposing that $|p_1|<\Lambda$. The effective potential
$V_0(\sigma)$, as a whole, must be a finite quantity at
$\Lambda\to\infty$. So the UV divergence of the integral term in
(\ref{A1}) must be compensated by the term with the bare coupling
constant $G$, which, of course, has to be a $\Lambda$-dependent
quantity. To find an appropriate expression for $G$, let us recall
that in the vacuum the chiral symmetry is necessarily broken down in
the framework of the model (\ref{1}), and quarks acquire a nonzero
dynamical mass $M_0$ which is a nontrivial solution of the gap
equation $V_0^{'}(\sigma)=0$. Taking into account this circumstance,
one can immediately obtain from the gap equation the expression
(\ref{16}) for the bare coupling $G$. Substituting it in (\ref{A1}),
it is possible to get for $\Lambda >>M_0$ the effective potential
(\ref{17}) of the initial model in the vacuum and in the $R^1\times
R^1$ spacetime. (More details about the above renormalization
procedure for $V_0(\sigma)$ are presented, e.g., in
\cite{kgk,vshivtsev}).

Now let us find the effective potential $V_{L}(\sigma)$ of the model
(\ref{1}), when the spacetime has a nontrivial topology of the form
$R^1\times S^1$ and quark fields obey the most general boundary
conditions (\ref{18}). In this case one can start from the equation
(\ref{A1}), in which it is necessary to perform the euclidian
rotation ($p_0\to i p_0$) and then use the transformations 
according to the rule (\ref{19}). As a result, we have
\begin{eqnarray}
V_L(\sigma)=\frac{\sigma^2}{4G}-\frac
2L\sum_{n=-\infty}^{\infty}
\int \frac{dp_0}{2\pi}\ln \left
[p_0^2+\sigma^2+\frac{\pi^2}{L^2}(2n+\alpha)^2\right ].
\label{A2}
\end{eqnarray}
Let us next use in (\ref{A2}) the formula $\ln a=-\int_0^\infty
\frac{ds}s e^{-as}$,
which is valid up to an infinite constant independent of the
parameter $a$, as well as the Poisson sum formula \cite{poisson}
\begin{eqnarray}
\sum_{n=-\infty}^{\infty}
e^{- s\frac{\pi^2}{L^2}(2n+\alpha)^2}=
\frac{L}{2\pi}\sqrt{\frac\pi s}\sum_{n=-\infty}^{\infty}
e^{-\frac{n^2L^2}{4s}}e^{in\pi\alpha}=
\frac{L}{2\pi}\sqrt{\frac\pi s}\biggl\{1+2\sum_{n=1}^{\infty}
e^{-\frac{n^2L^2}{4s}}\cos( n\pi\alpha)\biggr\}.
\label{A3}
\end{eqnarray}
After integration over $p_0$, one can easily find
\begin{eqnarray}
V_L(\sigma)=V_0(\sigma)+\frac{1}{\pi}\sum_
{n=1}^{\infty}\int_0^\infty\frac{ds}{s^2}
e^{-\sigma^2s-\frac{n^2L^2}{4s}}\cos( n\pi\alpha),
\label{A4}
\end{eqnarray}
where
\begin{eqnarray}
V_0(\sigma)=\frac{\sigma^2}{4G}+\frac{1}{2\pi}\int_0^\infty
\frac{ds}{s^2} e^{-\sigma^2s}
\label{A5}
\end{eqnarray}
is another, equivalent, expression for the effective potential
(\ref{A1}). (It is clear from general considerations that at
$L\to\infty$ the effective potential $V_L(\sigma)$ must coincide
with the effective potential in the $R^1\times R^1$ spacetime, i.e.
with $V_0(\sigma)$ given in (\ref{A1}). Looking at the formula
(\ref{A4}) (or (\ref{A6}) below) at $L\to\infty$, one can easily
obtain in the right hand side the expression appearing in the right
hand side of formula (\ref{A5}). So, due to the
above reason, it is the function $V_0(\sigma)$.) Moreover, in the
following we will use for $V_0(\sigma)$ its renormalized expression
(\ref{17}). Let us integrate in (\ref{A4}) over $s$ using the well
known relation \cite{prudnikov}
\begin{eqnarray}
\int\limits^{\infty}_{0}
dss^{n -1} e^{-\frac{A}{s}-Bs}=
2\left (\frac AB \right )^{n/2}K_n(2\sqrt{AB}),
\end{eqnarray}
where  $K_n(z)=K_{-n}(z)$ is the Macdonald function \cite{poisson}.
Then,
\begin{eqnarray}
V_L(\sigma)=V_0(\sigma)+\frac{4\sigma}{\pi}
\sum_{n=1}^{\infty}\frac{\cos( n\pi\alpha)}{nL}K_1(nL\sigma).
\label{A6}
\end{eqnarray}
Due to the relation $z\frac{d}{dz}K_1(z)=-K_1(z)-zK_0(z)$, it follows
from (\ref{A6}) that
\begin{eqnarray}
\frac{\partial}{\partial\sigma}V_L(\sigma)=\frac{\partial}{
\partial\sigma}V_0(\sigma)-\frac{4\sigma}{\pi}
\sum_{n=1}^{\infty}\cos( n\pi\alpha)K_0(nL\sigma).
\label{A7}
\end{eqnarray}
The series in (\ref{A7}) can be modified appropriately
\cite{prudnikov}, so (here
we use the effective potential $V_0(\sigma)$ in its renormalized
form (\ref{17}))
\begin{eqnarray}
\frac{\partial}{\partial\sigma}V_L(\sigma)&=&-\frac{2\sigma}{\pi}
\ln\left (\frac{M_0L}{4\pi}\right
)-\frac{2\sigma\gamma}{\pi}-\frac{2\sigma}{\sqrt{\sigma^2L^2+\pi^2
\alpha^2}}\nonumber\\
&-&2\sigma\sum_{n=1}^{\infty}\left
[\frac{1}{\sqrt{\pi^2(2n+\alpha)^2+L^2\sigma^2}}-\frac{1}{2n\pi}
\right ]-2\sigma\sum_{n=1}^{\infty}\left
[\frac{1}{\sqrt{\pi^2(2n-\alpha)^2+L^2\sigma^2}}-\frac{1}{2n\pi}
\right ],
\label{A8}
\end{eqnarray}
where $\gamma=0.577...$ is the Euler constant. Integrating both sides
of (\ref{A8}) over $\sigma$, we obtain the final expression for the
effective potential $V_L(\sigma)$ of the initial GN model in the
vacuum, when the space coordinate is compactified
\begin{eqnarray}
V_L(\sigma)&-&V_L(0)=-\frac{\sigma^2} {\pi} \ln\left
(\frac{M_0L}{4\pi}\right )-
\frac{\sigma^2\gamma}{\pi}-\frac{2}{L^2}\sqrt{\sigma^2L^2+\pi^2
\alpha^2}+\frac{2\pi\alpha}{L^2}\nonumber\\
&-&\frac{2}{L^2}\sum_{n=1}^{\infty}\left
[\sqrt{\pi^2(2n+\alpha)^2+L^2\sigma^2}
+\sqrt{\pi^2(2n-\alpha)^2+L^2\sigma^2}-4n\pi-\frac{\sigma^2L^2}{2n\pi
}\right ].
\label{A9}
\end{eqnarray}
In spite of the fact that (\ref{A9}) is valid for arbitrary
$0\le\alpha<2$, we find it more convenient to have another
(equivalent) expression for the function $V_L(\sigma)$ at $\alpha=1$.
In this case one can start again from the relation (\ref{A8}), in
which in the second series it is necessary to shift the summation
index, i.e. $n\to n+1$. Then, manipulating with convergent infinite
sums, we obtain
\begin{eqnarray}
\frac{\partial}{\partial\sigma}V^{\alpha=1}_L(\sigma)&=&-\frac{2
\sigma}{\pi}\ln\left (\frac{M_0L}{\pi}\right
)-\frac{2\sigma\gamma}{\pi}-
4\sigma\sum_{n=0}^{\infty}\left
[\frac{1}{\sqrt{\pi^2(2n+1)^2+L^2\sigma^2}}-\frac{1}{(2n+1)\pi}
\right ].
\label{A10}
\end{eqnarray}
Now, in order to get the effective potential, one should integrate
both sides of this relation over $\sigma$:
\begin{eqnarray}
V^{\alpha=1}_L(\sigma)&-&V^{\alpha=1}_L(0)=
\frac{\sigma^2}{\pi}
\ln\left (\frac{\pi}{M_0L}\right )-\frac{\sigma^2\gamma}{\pi}
\nonumber\\&-&\frac{4}{L^2}\sum_{n=0}^{\infty}\left
[\sqrt{\pi^2(2n+1)^2+L^2\sigma^2}
-(2n+1)\pi-\frac{\sigma^2L^2}{2(2n+1)\pi}\right ].
\label{A11}
\end{eqnarray}

Finally, a few words about the phase structure of the model at finite
$L$. It is clear that if the antiperiodic boundary conditions are
imposed on the quark fields, i.e. $\alpha=1$ in (\ref{18}), the
parameter $L$ plays the role of the inverse temperature in the
ordinary GN model, but only in the vacuum (with zero chemical
potentials). In the last case the critical properties of the GN model
are well understood (see, e.g., in \cite{jacobs}), so, by analogy,
we can conclude that at $L<L_c=\frac{\pi}{M_0}e^{-\gamma}$ the
global minimum of the effective potential (\ref{A11}) lies at the
point $\sigma=0$. In this case the chiral symmetry of the initial GN
(\ref{1}) model is not broken. In contrast, at $L>L_c$ the effective
potential (\ref{A11}) has a nontrivial global minimum point. As a
result, chiral symmetry is spontaneously broken down at
sufficiently high $L$.

The situation is however quite different at periodic boundary
conditions, i.e. if $\alpha=0$ in (\ref{18}). In this case the global
minimum point of the corresponding effective potential (\ref{A9})
lies outside the point $\sigma=0$ for all $L\ne 0$. Indeed, it is
clear from (\ref{A8}) that the derivative of this function is
negative at sufficiently small values of $\sigma$, so there is always
a local maximum of the effective potential at the point $\sigma=0$,
and the chirally broken phase is realized in the model for arbitrary
$L\ne 0$.

\end{document}